# The Micromechanical Measurement of Photothermal Expansion.

## Abstract

In the present work I analyze the factors that affect signal generation in the measurement of photothermal expansion with an atomic force microscopy probe. I discuss the general properties of the signal transduction mechanism considering two forms of light excitation, progressive and impulsive, and I derive guidelines for the optimization of the measurement. I also provide a general framework to analyze the thermomechanical processes that underlie signal generation in terms of a thermodynamic engine cycle and I discuss the relationship between signal generation and efficiency of the mechanism. The work has been subdivided into sections for the sake of clarity.

# The Micromechanical Measurement of Photothermal Expansion. Part 1: Analysis of Parameters Affecting Signal Generation with Gradual Excitation

*Luca Quaroni*

*Department of Physical Chemistry and Electrochemistry, Faculty of Chemistry, Jagiellonian University, 30-387, Kraków, Poland e-mail: luca.quaroni@uj.edu.pl*

**Abstract of Part 1**

The measurement of photothermal expansion via the deflection of a cantilevered probe has attracted attention as a tool for spectroscopic analysis. The micromechanical detection scheme measures light absorption with spatial resolution better than allowed by diffraction and is the basis of nanoscale spectroscopic techniques. In the present section of the work, I provide a simple theoretical analysis of signal generation that relies on the mechanical description of constrained thermal expansion of a solid. I relate signal intensity to the thermomechanical properties of the sample and the probe, and their contact geometry for the specific case of gradual excitation. The resulting equations support the general applicability of the latter configuration in photothermal spectroscopy experiments and provide guidelines for optimization of the signal.

## 1.0 Introduction

The photothermal effect converts the light absorbed by a material into heat, leading to a local increase in temperature and pressure, and a corresponding volume expansion. The effect was first observed by Bell, when he reported the associated sound emission.[1] The magnitude of the photothermal expansion tracks the absorption coefficient of the material and is wavelength dependent, making it a useful probe of the spectroscopic properties of a sample. Over the years multiple measurement configurations have been conceived to exploit the spectroscopic capabilities of the effect.[2] Among them, the use of atomic force microscopy (AFM) instrumentation is receiving increasing attention. Expansion is monitored via the deflection of the cantilever of an AFM probe in contact with the sample, providing local information about light absorption.[3] The spatial resolution of the measurement is independent from the wavelength of the exciting radiation, creating the opportunity to avoid the limits set by diffraction. This characteristic is particularly valuable in the mid-infrared (mid-IR) spectral region, where the relatively long wavelength ($\lambda \sim 2.5 - 25$ μm) corresponds to a diffraction limited resolution of the order of a few micrometers.

The feasibility of this micromechanical detection scheme of photothermal expansion was demonstrated by Anderson with mid-IR light using continuous wave (CW) light sources.[3] The experiments relied on cantilever deflection to detect photothermal expansion in the nanometer range, easily accessible to standard AFM instrumentation (Figure 1). As in conventional AFM, movement of the AFM positioning laser on a four quadrant detector is used to quantify the deflection. They also provided a mechanically detected interferogram when using light modulated by the Michelson interferometer of a Fourier Transform Infrared (FTIR) optical bench, although no spectrum was provided by its Fourier-Transform. The method was later developed into a technique for spectroscopic analysis by Dazzi *et al.* using the higher power pulsed emission from a Free-Electron laser as a light source.[4] The AFM-based detection scheme was commercialized with benchtop pulsed laser sources, extending its accessibility to a larger pool of researchers. Different names are used to describe the various incarnations of this configuration, among which common ones are AFM-IR [4] and Photothermal-Induced Resonance (PTIR).[5] The former has been popularized with commercial products and has seen more widespread use, while the latter is restricted to the case of impulsive excitation. In the present work, the acronym AFM-IR will encompass all configurations.

AFM-IR measurements can be carried out as either spectromicroscopy or imaging experiments. In the former, the probe is kept stationary while the light source is scanned through the spectral range of interest, providing a local spectrum of the sample. In the imaging configuration the AFM tip is scanned over the sample while the light source is operated at a single wavelength. The outcome is a two-dimensional distribution of sample absorption, modulated by other factors, such as optical and thermoelastic properties. A multiplexing approach has also been described, whereby a broadband light source excites the sample and the cantilever response is analyzed by interferometry.[6]

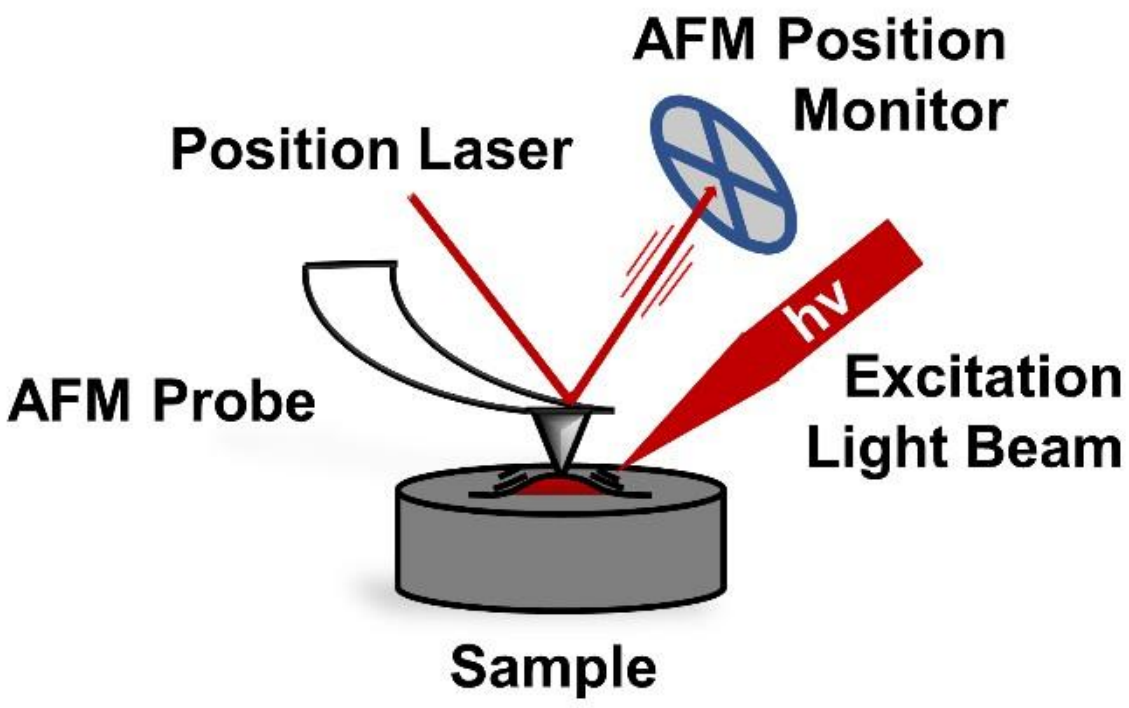

***Figure 1.*** *Micromechanical detection of photothermal expansion. The sample is mounted as for a conventional AFM experiment and irradiation occurs by focusing light at the tip-sample contact region. An absorbing material expands upon illumination and the expansion is recorded via the deflection of the cantilever and quantified by the movement of the AFM monitor laser on a four-quadrant detector.*

In this work, I distinguish between two different modes of excitation in AFM-IR experiments, which shall be called gradual and impulsive. Gradual excitation involves the use of a continuous wave (CW) laser, or a thermal source. In such configuration the sample is exposed to continuous illumination and the temperature of the sample increases progressively, until illumination is interrupted, for example by chopping the beam, or until the rate of heating is matched by the rate of cooling and a steady state is achieved.[7] It is taken to be a defining feature of gradual excitation that the cantilever deflection must track the expansion and contraction of the sample without loss of contact from the tip. For our purposes, a slow sinusoidal modulation of the light intensity is also considered a form of gradual excitation as long as the latter condition is met. In the case of impulsive excitation, the sample is rapidly heated by a laser pulse. The impulse excites resonant modes of the cantilever, giving rise to exponentially decaying oscillations, the amplitude of which is used as a proxy for light absorption.[4] In the most general case of impulsive excitation, the oscillations of the cantilever do not track the expansion and contraction of the sample in phase, except under resonant conditions.[8] Under these conditions, it is the complex movement of the cantilever resonating according to its normal modes that drives the oscillation of the laser on the four quadrant detector, instead of the simple deflection.

Despite the rising interest on applications of the technique, aspects of the signal generation mechanism are still unclear or neglected. Several theoretical treatments of signal generation have been developed to date, most of which have focused on impulsive excitation.[8, 9,10,11,12,13,14] Morozovska *et al.* provide a detailed treatment of the role of optical and thermoelastic properties in photothermal expansion using sinusoidal excitation.[15] However, it does not address the role of geometry and mechanical properties of the probe in the signal transduction train. It is the purpose of the present treatment to fill this gap. I will do so via a simple assessment of cantilever deflection and signal generation based on the classical description of constrained expansion of a heated solid. The analysis provides guidelines for optimizing AFM-IR experiments with gradual excitation and can also offer useful insight into the case of impulsive excitation.

## 2.0 Analysis and Discussion

The response of an AFM probe deflected by photothermal expansion under conditions of gradual excitation will be described using the classical treatment of a solid undergoing constrained thermal expansion. It is assumed that the temperature increase, $\Delta T$, is small, such that thermomechanical properties can be considered constant. The dependence of $\Delta T$ from the absorption coefficient and thermoelastic properties of the sample, and its wavelength dependence have been described elsewhere and will not be treated further. [15] The following analysis will be concerned with the thermoelastic properties of the sample and the probe, as summarized by Figure 2, showing the tip - sample contact region for two possible geometries. In the simplest case a flat tip, either a cone or a truncated cylinder, is in contact with a flat sample surface (panels A, C, E). In a more complex case, closer to the common experimental situation encountered in AFM-IR, a curved tip is in contact with a locally flat sample (panels B, D, F).

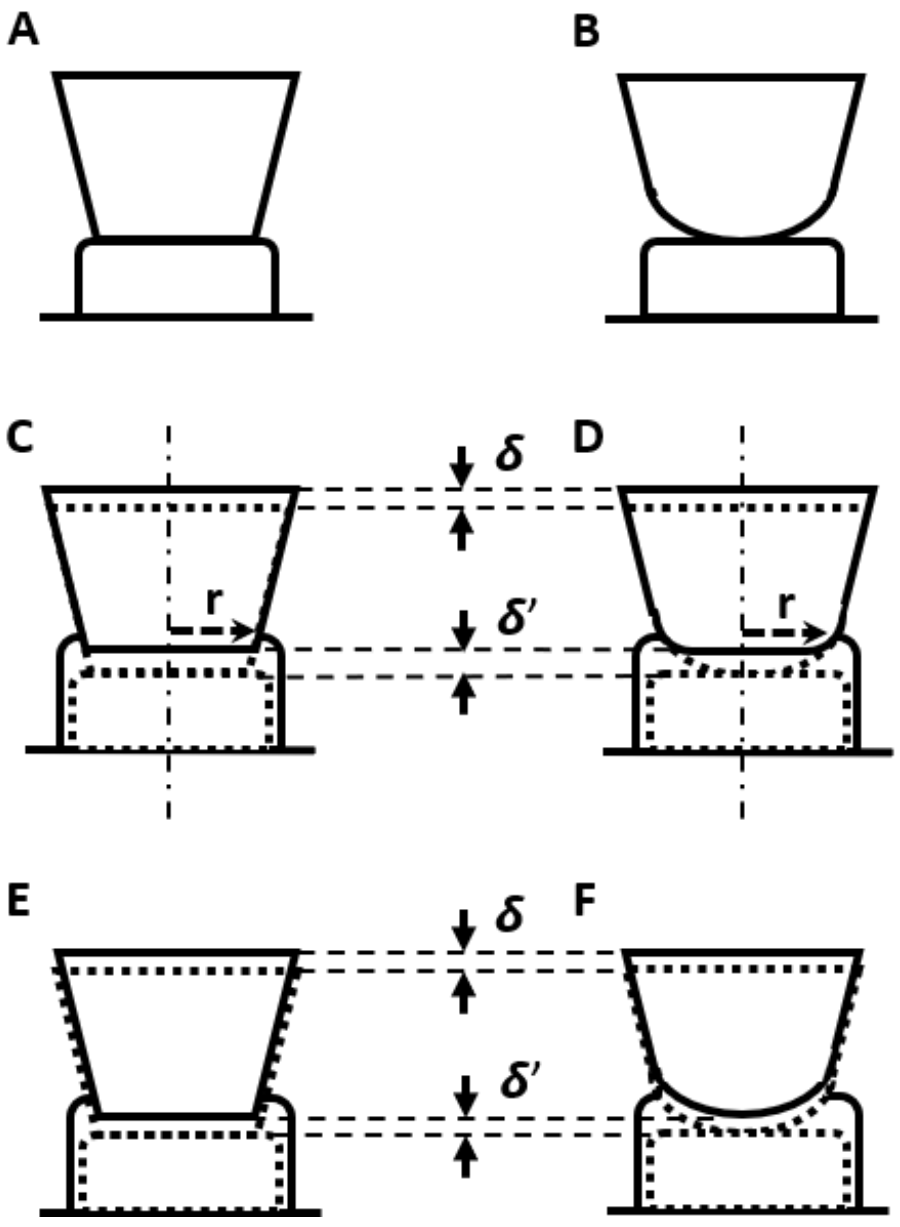

***Figure 2.*** *The tip-sample contact region in an AFM-IR experiment for the case of a flat tip (Panels A, C, E) or a curved tip (Panels B, D, F). A, B – resting position of the tip in contact with the sample before excitation. C, D – Photothermal expansion of the sample when tip and sample have similar rigidity (i.e., the Young moduli are comparable). The expansion causes compression of both tip and sample at the contact region, defining a contact surface of radius r. The contact is displaced by δ' from the initial position, while the back of the tip, at the cantilever contact, is displaced by δ. In general, δ ≠ δ'. E, F - Photothermal expansion of the sample when the sample is much more compressible than the tip (i.e., the Young modulus of the sample is much smaller than that of the tip). The expansion causes compression of the sample at the contact region, defining a contact surface of radius r. The displacement of the contact surface is the same as the back movement of the tip, δ = δ'.*

Photothermal heating of the sample causes its expansion and an increase in the pressure applied at the tip-sample contact. In the general case, the Young moduli of sample and tip are compatible and compression of both the sample and the tip occurs, together with upward tip displacement, δ (Panels C and D). The compression also causes an upward displacement of the tip-sample contact, δ', with δ ≠ δ. Assuming an axisymmetric tip, the projection of the contact surface is a disk of radius r. In the commonly encountered experimental situation where the material of the tip is much less compressible than the sample, the compression of the tip is negligible, and δ = δ' (Panels E and F). The displacement δ causes a proportional tilt the cantilever of the probe, moving the beam across the four-quadrant detector and generating the AFM-IR signal (Figure 1). Maximizing the AFM-IR signal corresponds to maximizing the displacement, δ, which is equivalent to maximizing the deflection of the cantilever.

For a given temperature increase, the maximum deflection is determined by balancing the force from the expansion of the working medium, $F_e$, with the compressive force exerted by the cantilever, $F_c$. $F_e$ is calculated as the force exerted by a compressible medium undergoing constrained expansion and is given by Equation (1).

$$F_e = E^* \alpha \Delta T A \qquad (1)$$

$E^*$ is the reduced Young modulus. α is the thermal expansion coefficient of the sample. ΔT is the temperature difference. A is the contact area between tip and sample. If the sample is larger than the tip base, as in Figure 2, A corresponds to the area of the tip base. In the case of a sample smaller than the tip base, A can be approximated by the sample surface. Equation 1 assumes a small compression, such that shear stress at the edges of the compression area is negligible for the case of a sample larger than the tip.

The deflection of the cantilever under the action of this force, is described by Hook's Law for the simple case of a harmonic spring (Equation 2), where K is the spring constant for the cantilever.

$$F_c = - \delta K \qquad (2)$$

If a setpoint force is being applied, the deflection of the cantilever will be different from zero in the absence of photothermal expansion. In such case δ is taken to be the total deflection, i.e. the sum of the setpoint deflection and the additional deflection caused by sample expansion. The maximum deflection, $\delta_{max}$, corresponding to a given temperature increase ΔT is obtained when $F_e = - F_c$ and is given by Equation 3.

$$\delta_{max} = (E^* \alpha \Delta T A)/K \qquad (3)$$

For the specific cases of a tip in the shape of a cylinder or a truncated cone, A is the area of the cylinder base or of the cone cross section respectively. For a spherical tip, with local radius r, we approximate the contact area to the projection of the tip on the sample surface (Equation 4). The approximation is acceptable for the common case of a hard tip (e.g., silicon) in contact with a soft material. [a]

$$\delta_{max} = (E^* \alpha \Delta T \pi r^2)/K \qquad (4)$$

The reduced modulus is defined by Equation (5).

$$E^{*\,-1} = (1-\sigma^2)/E + (1-\sigma'^2)/E' \quad (5)$$

E and E' are the Young modulus of the sample and of the tip respectively, whereas σ and σ' are the Poisson ratio of the sample and of the tip. If the modulus of the sample is much lower than the modulus of the tip (the common case of a hard tip and a soft sample), Equation (5) simplifies to Equation (6).

$$E^* = E/(1-\sigma^2) \quad (6)$$

Equations (3) and (4) express the dependence of $\delta_{max}$ from the thermomechanical and geometric properties of the sample and the probe. Materials with larger thermal expansion coefficient, α, deliver stronger signal. Stiffer materials, with larger values of E, deliver stronger signal. Using tips with a larger radius or larger contact surface also provides greater deflection. Maximum deflection decreases with increasing K, showing that softer cantilevers provide higher sensitivity. An important feature is the interplay of the elastic properties of both tip and sample, including Young moduli and Poisson coefficients, which jointly contribute to the signal. In soft matter experiments the elastic properties of the sample are the dominant contribution to E*. However, this is no longer the case when the sample is a hard material. In such case, the joint contribution of tip and sample needs to be evaluated if optimal response is desired. Several of these conclusions are qualitatively consistent with results from the analysis of impulsive excitation experiments, although from the quantitative point of view the dependence of the force from thermoelastic and geometric parameters of tip and sample is different. [9,11,12] However, contrasting conclusions are obtained from the analysis by Dazzi *et al.*, which provides an opposite dependence of signal intensity from the value of K. [14]

[a] To obtain a more accurate quantification of the contact surface, particularly when tip and sample have comparable elastic moduli, detailed models of contact mechanics need to be used. However, the general conclusions of the present work would still be valid, and this task is left for future work.

One important result of the present analysis is the rapid increase of the signal provided by larger tips and an expanded tip-sample contact. Wider tips generally deliver poorer spatial resolution in scanning probe measurements, thus highlighting the existence of a trade-off between the resolution and the sensitivity of an AFM-IR measurement. This conclusion can better inform experimental practice, where tip selection is commonly performed with a view on its contribution to resolution, whereas other aspects are commonly neglected.

Table I summarizes this discussion and can be used both as a guideline for the optimization of AFM-IR experiments and as a reference to test the model described in this work.

***Table I. Effect of Sample and Probe properties on the AFM-IR signal.*** *The "+" sign indicates quantities that correlates positively with an increased signal, while the "-" sign indicates quantities that correlate negatively. K: Cantilever Force Constant. A: Tip/Sample contact surface. E, E': Young Moduli of Sample and Tip. α: Thermal Expansion Coefficient of Sample. σ, σ': Poisson Ratios of Tip and Sample. r: Radius of tip/sample contact surface. ΔT: Photothermal Induced Temperature Increase.*

| Quantity | Predicted Correlation with Deflection |
|---|---|
| K | - |
| A, r | + |
| E, E' | + |
| α | + |
| σ, σ' | + |
| ΔT | + |

The present treatment provides the maximum deflection for a unit temperature increase. No attempt is made here to relate the increase in temperature to the spectroscopic properties of the sample, nor to address the effect of heat transfer to the surrounding environment. A theoretical treatment of the latter aspect has been carried out by Morozovska *et al.*[15] Other theoretical treatments of heat transfer in impulsive experiments have been provided by several groups,[9,12,13] together with an experimental investigation of its interplay with resolution [16] and the reader interested in this contribution is directed to these works.

In the implementation of an experiment with gradual excitation, ΔT increases over time when the sample is irradiated at constant power, as reported experimentally. Extended irradiation leads asymptotically to a thermal steady state, as ΔT plateaus to a stable value, which is determined by thermal exchanges between the medium and the environment. If the excitation beam is modulated by a chopper, a see-saw pattern of expansion and contraction is expected.[3]

Expansion amplitudes of the order of the nanometer have been reported using gradual excitation with mid-infrared light.[3] We can assess the validity of Equation (3) by comparing its predictions with this benchmark. Let us consider a sample of polystyrene (PS), for which $E = 3x10^9$ Pa, $\sigma = 0.35$ and the linear thermal expansion coefficient $\alpha = 7x10^{-5}\ K^{-1}$.[17] We irradiate the sample with light at the frequency of the absorption bands of PS, producing a ΔT = 1 K, a realistic value for the T increase in a polymer sample irradiated with a light source of a few 100's µW (such as from a thermal source for FTIR spectroscopy or a laser). We monitor the expansion using a contact mode AFM probe with K = 1 N/m and the radius of the contact area r = 50 nm. Equation (3) provides a value of maximum deflection $\delta_{max} = 1.9$ nm, close to the value of 1.5 nm reported for this system, [3] as allowed by the wide variability in published values of E and α. While this is only an approximate verification, it shows that Equation (3) produces values that are consistent with available experimental results.

## 3.0 Conclusions

I have used the classical treatment of constrained thermal expansion to provide a theoretical description of signal generation in AFM-IR experiments with gradual excitation. The analysis provides the dependence of the signal from the thermomechanical and geometrical properties of tip and sample. Less compressible materials provide higher signal for an equivalent temperature increase. Most importantly, I note a linear dependence of the signal from the contact area of tip and sample, which implies a trade-off between resolution and signal intensity.

The analysis supports the general viability of an AFM-IR configuration with gradual excitation for spectroscopic applications. Samples accessible by this configuration are relevant for a variety of disciplines, ranging from polymer science and surface chemistry, to electrochemistry, biochemistry, and biology, including materials of technological interest. Interest in this mechanism arose because of its spectroscopic applications, as a variation of the AFM-IR experiment; nonetheless, it is of more general utility, because it allows quantitative measurement of multiple properties (as defined by Equation (3)) on microscopic samples, including temperature variations.

While the present calculations are limited to the case of gradual excitation, some of the conclusions can be qualitatively extended to impulsive experiments. For such case, the amplitude of the signal is predicted to increase with the amount of energy transferred to the cantilever by the impulse of the expansion force. Therefore, quantities affecting $F_e$ are also expected to similarly affect the impulsive signal, as long as no other forces (e.g., electrostatic, magnetic photoacoustic, photoinduced) are involved in signal generation.[18] Provided that the latter condition is met, the guidelines of Table I can be considered of general validity in the design of AFM-IR experiments.

## Conflict of Interest

The author declares no conflict of interest.

# The Micromechanical Measurement of Photothermal Expansion. Part 2: Description of the Measurement via an Engine Cycle

*Luca Quaroni*

*Department of Physical Chemistry and Electrochemistry, Faculty of Chemistry, Jagiellonian University, 30-387, Kraków, Poland e-mail: luca.quaroni@uj.edu.pl*

**Abstract of Part 2**

Photothermal expansion can be measured with high spatial resolution via a micromechanical detection scheme based on the deflection of a cantilevered probe for atomic force microscopy. In this section of the work I provide a conceptual framework to analyze the processes that underlie the measurement by describing them in terms of a thermodynamic engine cycle. I discuss the general properties of this photothermal micromechanical engine for the two general cases of gradual excitation and impulsive excitation. For the case of gradual excitation, the engine can be described by a thermodynamic cycle involving quasi-static processes. Nonetheless, this engine is intrinsically inefficient when assessed in its capacity to convert energy into work, and most or all of the energy derived from light absorption is dispersed into the environment as heat. I also argue that thermodynamic efficiency is not a critical performance parameter when considering the main purpose of this mechanism, which is analytical. For the case of impulsive excitation, I note that the irreversibility of several of the processes and the complexity of tip-sample interactions limit the applicability of a thermodynamic cycle that relies on quasi-static processes. Nonetheless, the mechanism appears capable of producing usable work with cyclic operation and can be used to this purpose, in addition to its conventional analytical application.

## 1.0 Introduction

The photothermal effect converts light absorbed by a material into heat, increasing temperature and pressure in the absorbing material, causing its expansion. The expansion is wavelength dependent and tracks the expansion coefficient of the material, leading to its application in spectroscopic analysis.[1] The use of atomic force microscopy (AFM) instrumentation for quantitatively monitoring the expansion is the most recent development and relies on the deflection of the cantilever of an AFM probe in contact with the sample.[2] In this configuration, the spatial resolution of the measurement does not depend on the wavelength of absorbed light, thus bypassing restrictions due to the optical diffraction limit.[3] Applications are particularly valuable in the extensively used mid-infrared (mid-IR) spectral region ($\lambda \sim 2.5 - 25$ µm), where resolution of 1/100 $\lambda$ or better can be achieved. The detection of photothermal expansion from mid-IR light using the deflection of an AFM cantilever was demonstrated by Anderson with standard instrumentation for AFM and IR spectroscopy, using both a thermal source and a continuous wave laser.[2] The capability to perform spectroscopic analysis was later developed by Dazzi *et al.* using the pulsed emission from a Free-Electron laser and from a benchtop laser as light sources.[4] The technique was termed Photothermal Induced Resonance (PTIR) when implemented with pulsed lasers, because it relies on the excitation of resonant oscillations of the cantilever for signal generation, as opposed to simple cantilever deflection. Following the commercialization of instruments for this measurement, the more general name AFM-IR has been introduced. In the present work, AFM-IR will be used as a general moniker to encompass all configurations.

AFM-IR measurements can be distinguished as experiments employing gradual excitation or impulsive excitation. In the case of gradual excitation, the sample is exposed to illumination that is constant or slowly modulated, over times that are much longer than the rate of thermal relaxation, using a continuous wave (CW) laser or a thermal source. Under these conditions, the temperature increases progressively until a steady state is achieved, or until the beam is interrupted or attenuated (e.g., by chopping), and the deflection of the probe cantilever tracks the expansion and contraction of the sample without loss of contact between tip and sample. [5] A sinusoidal modulation of the light intensity qualifies as gradual excitation if the latter condition is met. In the case of impulsive excitation, the sample is rapidly heated by a short laser pulse. The rapid impulse from the expanding sample excites resonant modes of the cantilever, giving rise to exponentially decaying oscillations. The amplitude of the oscillations, either measured directly or extracted from the resonance spectrum of the oscillating cantilever, is used for assessing the amount of absorbed light.[4] The oscillations are out of phase with the expansion and contraction cycles of the sample, except in the case of resonant excitation.[6]

Some basic aspects of signal generation in AFM-IR remain unclear, in part because of the complexity of the micromechanical signal transduction mechanism, which relies on the interplay of spectroscopic, thermal, and mechanical properties of the sample and of the probe. It is the purpose of the present analysis to expand the theoretical modelling of such mechanism.

## 2.0 Analysis and Discussion

### *2.1 The Micromechanical Detection Scheme of Photothermal Expansion is Described as a Photothermal Engine.*

It is appreciated that the scanning probe detection mechanism introduces a dependence of the signal from tip-sample contact mechanics, [6,7,8,9,10] and from heat-transfer processes within the sample and between tip and sample. [7,11,12,13] To gain further insight into the underlying processes, I am proposing to complement existing descriptions, which rely on theories of heat conductivity and thermoelasticity, with a treatment based on classical thermodynamics. The starting point of the discussion is the observation that, in thermodynamic terms, the processes involved in an AFM-IR experiment can be collectively described by the cycle of a microscopic engine, a device that converts energy into work. Energy from absorbed photons heats the sample which, by expanding, delivers the work necessary to deflect the cantilever of the AFM probe. When the sample cools back to ambient temperature, heat is released into the environment as the cantilever returns to its resting position. The cycle is analogous to that of a classical heat engine, with two main differences, one being that the working medium (i.e., the sample) is a solid instead of a gas and the other being that the expansion stage is used to increase the potential energy of the cantilever. The differences have important consequences on the energy balance of the cycle, as discussed later.

Figure 1A provides a graphical representation of the basic micromechanical device. The sample is mounted as in a conventional AFM experiment and irradiation occurs by focusing light at the tip-sample contact region. The photothermal expansion is recorded via the deflection of the cantilever and quantified by the movement of the AFM monitor laser on a four-quadrant detector. The deflection is generally assumed to be proportional to the local absorption coefficient of the sample at the wavelength of irradiation and is used for quantitative spectroscopic analysis.[4]

Figure 1B shows the transformations of the sample over the full cycle of excitation and relaxation. The sample acts as the working medium of the engine, and the two terms are used interchangeably in this work. The full cycle is ideally broken down into quasi-static reversible transformations (the strokes of the engine), connecting four states described by the set of thermodynamic variables, pressure (P), volume (V) and temperature (T). State **1** is the resting state. In the first stroke, absorbed light delivers energy to the medium, simultaneously increasing both T and P, and bringing it to state **2**. In the second stroke, the sample expands, reaching state **3**. In the transition from state **2** to state **3**, expansion acts against the force exerted by the cantilever and causes its deflection. In the third stroke, the expanded sample cools down to the initial temperature leading to state **4**. In the fourth and last stroke, the sample is compressed back to its initial volume, corresponding to the initial state **1**.

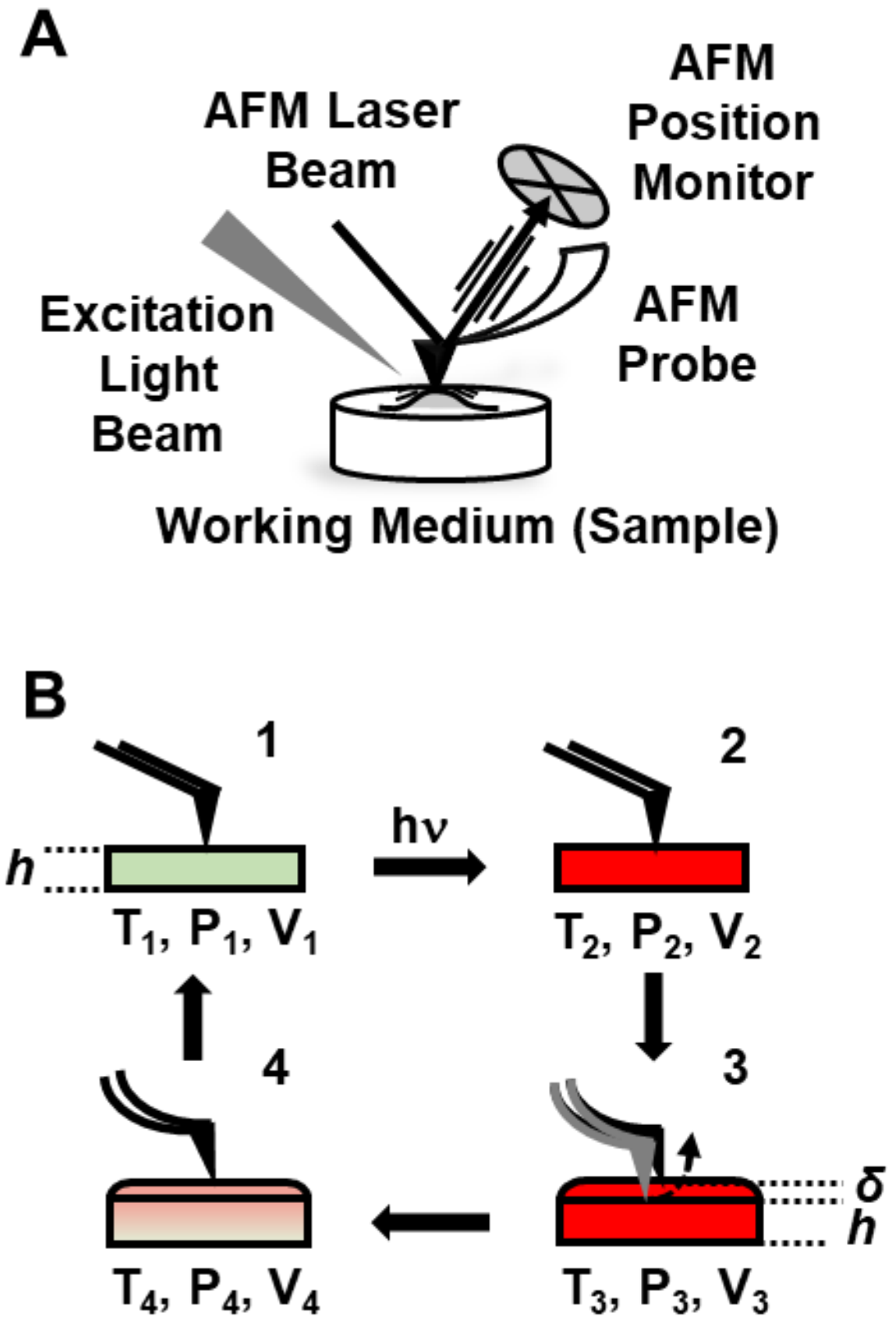

***Figure 1. Photothermal signal generation as an engine cycle**. **A.** AFM measurement of photothermal expansion. Irradiation of a light absorbing material causes localized expansion. The displacement of an AFM probe located at the absorbing region is used to monitor the expansion. **B.** Elementary states and transformations that are cycled through in the course of a measurement. The combined transformations correspond to an engine cycle. $T_n$, $P_n$, $V_n$: respectively, temperature, pressure, and volume of state n. d: sample thickness. δ: cantilever deflection. h: sample thickness. Red color graphically indicates higher temperature. **1**- Initial*

*state. **2**- After the temperature increase caused by light absorption. **3**- After expansion from state **2**. **4**- After cooling of state **3**.*

In this analysis I assume that the quantum yield for the conversion of absorbed light into heat is unity. I neglect any other radiative and non-radiative channels that may lead to dissipation of energy from absorbed photons, such as photochemical processes, fluorescence, phosphorescence. I also assume that the only force that acts on the AFM probe is the mechanical one arising from the action of the expanding medium. Therefore I also neglect the possibility of other interactions, such as those of electromagnetic or photoinduced origin, or short distance interactions arising from Van der Waals forces.

*2.2 Gradual Excitation.*

It is integral to the definition of gradual excitation that the sample expands while retaining contact with the tip of the probe. The slow operation ensures that the approximation of quasistatic reversibility can be applied to the cycle. Under conditions of gradual excitation, heating and expansion can be considered to occur simultaneously, providing a simpler cycle compared to the general one shown in Figure 1B. Strokes **1 – 2** and **2 – 3** from Figure 1B can be represented as a single transformation, bringing the sample from an initial state **1** represented by $T_1$, $P_1$, $V_1$ to a state **3** represented by $T_3$, $P_3$, $V_3$, with $T_3 > T_1$, $P_3 > P_1$, $V_3 > V_1$. Similarly cooling and compression are simultaneous and bring the system back from state **3** to state **1**. Figure 2 shows the conversion between these two states as a path in a Pressure-Volume (PV) diagram (solid line). The path extends between the two hyperboles that represent isothermal transformation paths at temperatures $T_1$ and $T_2$ (dashed lines). During the heating and expansion stroke, from **1** to **3**, heat is absorbed by the working medium in the form or energy from absorbed light (**$Q_{in}$**) and work is executed by the expanding medium on the AFM probe (**$W_{out}$**). During the cooling and compression stroke, from **3** to **1**, heat is released by the working medium to the environment (**$Q_{out}$**) and work is executed by the AFM probe on the contracting medium (**$W_{in}$**) (Figure 2A). The increase in volume of the medium triggers signal transduction by deflecting the cantilever (δ). Assuming an isotropic expansion, the deflection increases monotonically with sample expansion according to $\delta \propto (V_2 - V_1)^{1/3}$.

The harmonic response of the cantilever sets this mechanism apart from that of most conventional heat engines. Work output from the expanding medium (**$W_{out}$**) is converted into potential energy by the bending of the cantilever, instead of being transduced farther downstream. As the medium cools down and contracts, the cantilever fully returns the stored potential energy in the form of compression work (**$W_{in}$**) performed on the medium. If the tip maintains contact with the sample throughout the cycle, stroke **3 – 1** is the exact reverse of **1 – 3** and total work output ($W_{tot} = W_{out} - W_{in}$) from the engine is null when the cantilever has returned to its resting position. The energy provided by absorbed light has been totally transferred to the environment as heat, corresponding to zero engine efficiency (η), with $\eta = W_{tot}/Q_{in}$. An engine operating according to of Figure 2A is not capable to convert heat into work with cyclical operation. Nonetheless, deflection of the cantilever takes place, and provides an AFM-IR signal, even in the absence of net $W_{tot}$. Figure 2B shows the deflection versus time plot for the case of a beam modulated by mechanical chopping, in agreement with the response experimentally reported by Anderson. [2]

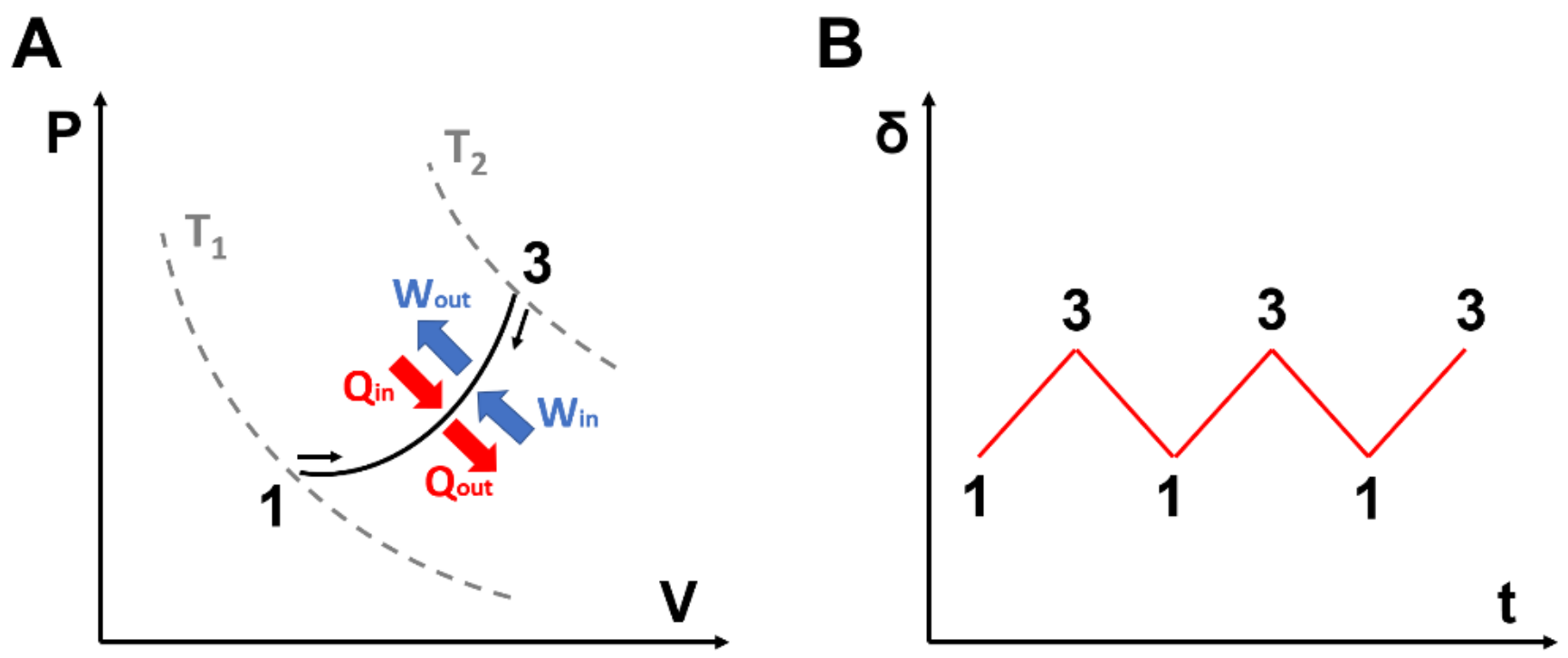

***Figure 2. The engine cycle under conditions of gradual excitation. A.*** *Engine cycle in a PV diagram. The medium absorbs light, progressing from state **1** to state **3**, with a corresponding increase in pressure, volume, and temperature. Interruption of illumination leads to cooling and decompression of the sample, returning to state **1** by reversing the expansion stroke. **B.** Deflection (δ) vs. time (t) diagram corresponding to the cyclical repetition of the two strokes in panel A.*

Neither work output nor efficiency directly relate to deflection and to the intensity of the signal, which is determined by the deflection of the cantilever. Therefore, the quantities that are normally used to characterize the performance of classical engines appear to have no direct consequence on the performance of the AFM-IR measurement under gradual excitation. Although this conclusion may appear to be a discrepancy, the discrepancy is only apparent. In contrast to a classical engine, which aims to convert heath into mechanical work, the purpose of the photothermal engine in AFM-IR is analytical. The movement of the engine in response to the presence of an absorbing sample is converted to a voltage difference at the four-quadrant detector, which in-turn provides information on the presence or absence of an absorbing material. In this context, the mechanism demonstrated by Anderson [2] acts as a machine that processes information, even while failing to produce useful work. Optimization of the signal from this mechanism can be performed according to the thermoelastic properties of tip and sample, and has been described elsewhere. [5]

*2.3 Impulsive Excitation.*

Impulsive excitation is the prevalent configuration in contemporary AFM-IR experiments, which rely on trains of laser pulses of duration ranging from nanoseconds to microseconds.[7] In this configuration the temperature of the sample increases rapidly, tracking the rise time of the pulse, then decreases exponentially after interruption of the pulse.[14,15], Impulsive expansion of the sample results in the rapid deflection of the cantilever and excitation of its resonant modes. Relaxation occurs via an exponential decay process with a ringdown pattern that releases energy to the surrounding environment.[4,16] Because of the timescales involved and the major contribution of dissipative processes, this mechanism is the least amenable to a description that relies on a thermodynamic cycle that approaches the reversible quasi-static limit. Nonetheless, some useful guidelines can be obtained for the case of a long pulse duration (i.e., the duration of the pulse is non-negligible compared to the time required for thermal exchange with the environment). In the case of an extended square pulse (black dashed line in Figure 3A), a rapid increase of the temperature of the sample is followed by an asymptotic convergence to a steady-state temperature (red solid line in Figure 3A).[17] After conclusion of the pulse, cooling of the medium occurs via an exponential process.

The rapid rise time of the pulse supports the assumption that most of the temperature increase occurs under isochoric conditions, while expansion occurs under conditions that are isothermal. The two latter transformations are represented in the PV diagram cycle of Figure 3B. Heating of the medium by light absorption occurs during the isochoric heating stroke **1 – 2** and during the thermostatic expansion stroke **2 – 3**. The two strokes are equivalent to strokes **1 – 2** and **2 – 3** of Figure 1B. Cooling and compression occur simultaneously during stroke **3 – 1** in an irreversible process (dashed line in Figure 3B).

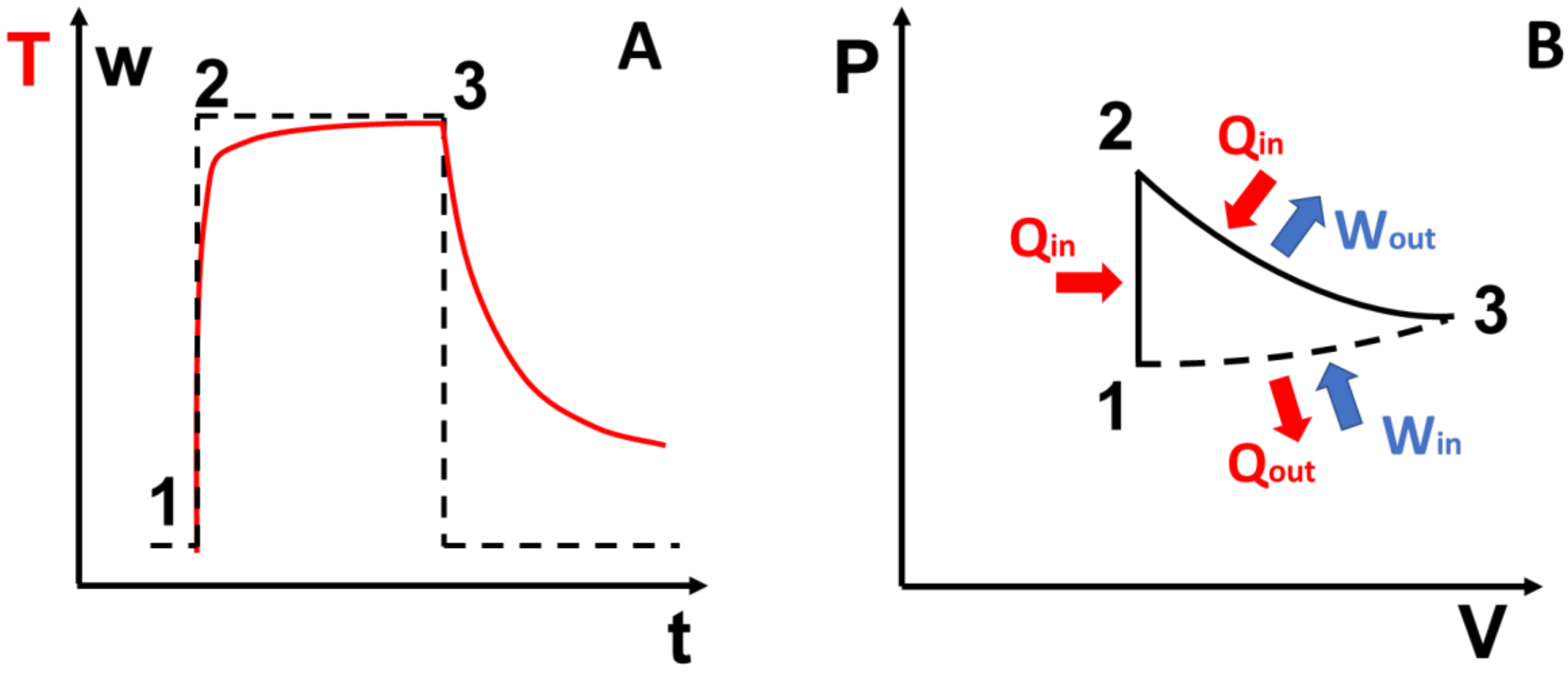

***Figure 3. Impulsive excitation and its thermodynamic cycle. A.*** *Pulse power (w, dotted line) and temperature increase (T, red line) within the working medium during a pulse. The temperature increases rapidly and approaches asymptotically the maximum value during the pulse. It decreases exponentially after the end of the pulse. The numbers correspond to those of the states in the next panel.* ***B.*** *Representation of the process in a PV diagram. Rapid sample heating by the pulse corresponds to an isochoric transformation from state* ***1*** *to state* ***2****. Sample expansion during irradiation corresponds to the transformation between* ***2*** *and* ***3****. Sample contraction and cooling correspond to the transformation between* ***3*** *and* ***1****.*

Energy is absorbed by the medium as photons (represented as an equivalent heat input $\mathbf{Q_{in}}$) during irradiation in strokes **1 – 2** and **2 – 3** and is released to the environment during cooling along **3 – 1** as heat ($\mathbf{Q_{out}}$). Work is executed by the medium on the cantilever during expansion **2 – 3** ($\mathbf{W_{out}}$) and is executed on the medium during compression and cooling **3 – 1** ($\mathbf{W_{in}}$). In contrast to the case of gradual excitation, the cycle in Figure 3B allows for non-zero values of $W_{tot}$ and $\eta$. Both these quantities are expected to be small in AFM-IR experiments, limited by the temperature increase, typically in the range of a few degrees Kelvin, and by the extent of the expansion stroke **2 – 3**, of the order of the nanometer.

The compression and cooling stroke **3 – 1** in Figure 3B is not amenable to a quasi-static approximation. In a spring-loaded engine with a gaseous medium, expansion increases the potential energy of the spring and compression completely returns this energy to the medium, providing no total work output. In contrast, using a solid working medium allows decoupling the contraction of the medium from the compression by the cantilever, leading to an irreversible transformation that does not satisfy the conditions for a quasi-static process. This point can be clarified by analyzing a limiting case via a *thought experiment*. Figure 4A shows a modified version of the cycle of Figure 1B. Expansion of the sample deflects the cantilever but, differently from Figure 1B, the cantilever is then locked in position by some mechanism that does not affect the energy balance of the system. While the medium cools and contracts, the cantilever retains the stored potential energy and loses contact with the medium. Once the medium has returned to the initial state, the retention mechanism is released, and the cantilever is allowed to return to its equilibrium position. The cantilever accelerates, as potential energy is converted into kinetic energy. Using a suitable mechanism, work can be extracted directly from the stored potential energy or from the kinetic energy of the moving tip. If the kinetic energy is not extracted, it is converted to oscillations of the cantilever following impact with the medium, and is later dissipated via available channels.

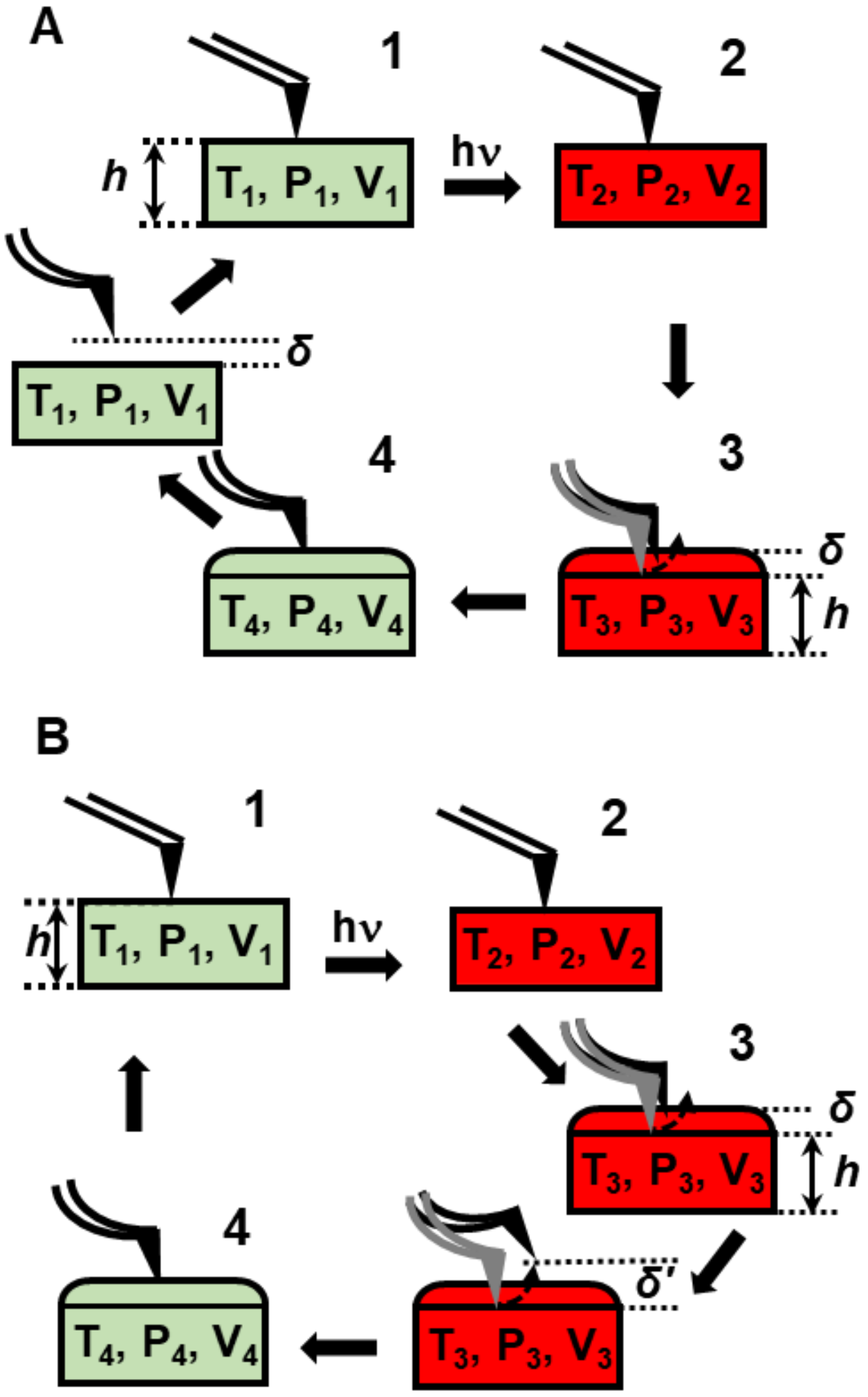

***Figure 4. Probe dynamics with impulsive excitation. A.*** *Loss of tip-sample contact during contraction: a thought experiment. Strokes 1-2, 2-3, and 3-4 are similar to the ones described in Figure 1 and expansion of the medium leads to deflection of the cantilever. Differently from Figure 1, between state 4 and state 1, the cantilever is locked in the deflected position while the medium cools and contracts, leading to loss of contact. Release of the cantilever converts*

*its potential energy into kinetic energy, triggering its oscillation, while the medium returns to the initial state.* ***B.*** *Loss of tip sample-contact during impulsive excitation. Strokes 1-2, 2-3, and 4-1 are similar to the ones described in Figure 1. Differently from Figure 1, the tip loses contact with the sample after expansion because of the impulse received from the rapidly expanding medium.*

In a real experiment, loss of synchrony between the contraction of the medium and the compression of the cantilever permits the acceleration of the tip. This is also true for the expansion stage, when triggered by a rapid short pulse. The impulse transmitted to the tip is capable of accelerating it and cause loss of contact with the medium (Figure 4B). It has been previously assumed, but never verified, that in impulsive AFM-IR experiments the probe retains contact with the sample during oscillation, allowing selective excitation of contact resonances of the cantilever. [7] I now argue that the tip can be displaced from the sample and excitation of non-contact resonances can take place. The measured oscillating decay results from overlapping contributions of contact and non-contact resonances. In the case of loss of tip-sample contact and of acceleration of the tip, the conditions for a quasi-static transformation do not apply, thus limiting the applicability of arguments of classical equilibrium thermodynamics.

The mechanics of impulsive energy transfer to the cantilever are beyond the scope of this work and will not be discussed further. Nonetheless, it is important to remark that displacement of the tip from the surface has important implications. If the distance between tip and sample deviates from zero over time, additional forces from tip-sample interactions can constrain cantilever deflection, affect damping, and contribute to the overall signal. Contributions to the total force can come from electrostatic [18] and magnetic interactions [19], including photoinduced forces. [20] The relative contribution of van der Waals forces [21] and of forces from photoacoustic expansion [22] would also change with any variation of tip-sample distance. As a result, the response of AFM-IR with pulsed excitation would not be a simple manifestation of the photothermal effect but the result of overlapping contributions from multiple processes.

## 3.0 Conclusions

I have described the detection scheme of the AFM-IR experiment in terms of a light-driven micromechanical engine, and I have proposed to analyze the properties of such mechanism within the framework of a thermodynamic cycle.

In the case of gradual excitation, the transformations of the cycle can be executed under approximately quasi-static conditions, and can be considered to be close to the reversibility limit. Differently from classical heat engines, this cycle does not produce any useful work on the environment and has zero efficiency. The lack of efficiency is not inconsistent with its analytical function, which the processing of information, rather than the conversion of heat into work.

I have commented on the possibility to extend a thermodynamic engine analysis to the case of impulsive excitation. For this case I note that the complexity of the mechanism and the irreversibility of some transformations limit the applicability of arguments based on equilibrium thermodynamics. Nonetheless, I observe that the impulsive mechanism appears capable of cyclical work production and can be used for this purpose, in addition to its analytical application.

I also remark that a description of the impulsive mechanism must account for the displacement of the probe from the sample, which is neglected in existing theoretical descriptions. Tip displacement from the surface introduces additional contributions from multiple interactions between tip and sample, which greatly increase the complexity of the analysis and limit our capability to develop a theoretical treatment. The possibility of displacement implies that AFMIR experiments with continuous and with impulsive excitation are fundamentally different, and only the former rely predominantly on the photothermal effect, while the latter are responsive to additional processes and interactions.

In conclusion, it must be stressed that the present discussion is based on classical thermodynamics and is valid as long as materials can be described by their bulk properties. This is commonly encountered with AFM-IR samples, which are generally thicker than 100 nm. However, AFM-IR experiments have also been extended to samples as thin as a molecular monolayer by using resonant mode excitation. Such measurements straddle into the realm of single molecule physics and of quantum mechanics and require models other than the macroscopic ones provided here for a satisfactory description.